\documentclass[prl,aps,twocolumn,showpacs]{revtex4}
\usepackage{bm}
\usepackage{graphicx}

\begin{document}

\title{Atomic spin decoherence near conducting and superconducting films}
\author{S.~Scheel}
\email{s.scheel@imperial.ac.uk}
\author{P.K.~Rekdal}
\author{P.L.~Knight}
\author{E.A.~Hinds}
\affiliation{Quantum Optics and Laser Science, Blackett Laboratory,
Imperial College London, Prince Consort Road, London SW7 2BW, United Kingdom}

\date{\today}

\begin{abstract}
We derive scaling laws for the spin decoherence of neutral atoms
trapped near conducting and superconducting plane surfaces. A new
result for thin films sheds light on the measurement of Y.J.~Lin,
I.~Teper, C.~Chin, and V.~Vuleti\'{c} [Phys. Rev. Lett.
\textbf{92}, 050404 (2004)]. Our calculation is based on a
quantum-theoretical treatment of electromagnetic radiation near
metallic bodies [P.K.~Rekdal, S.~Scheel, P.L.~Knight, and
E.A.~Hinds, Phys. Rev. A \textbf{70}, 013811 (2004)].  We show
that there is a critical atom-surface distance that maximizes the
spin relaxation rate and we show how this depends on the skin
depth and thickness of the metal surface. In the light of this
impedance-matching effect we discuss the spin relaxation to be
expected above a thin superconducting niobium layer.
\end{abstract}

\pacs{42.50.Ct, 34.50.Dy, 03.75.Be}

\maketitle

Trapped neutral atoms have intrinsically long coherence times,
making them suitable for many proposed applications based on
quantum state manipulation. These include interferometry
\cite{interferometry}, low-dimensional quantum gas studies
\cite{quantum gas}, and quantum information processing
\cite{Jaksch,Calarco,proposals}. The trapping structures required
for these applications typically have feature sizes on the micron
or sub-micron scale, sizes that are comparable with the atomic de
Broglie wavelength. The required trap frequencies are typically in
the 1\,kHz to 1\,MHz range, this being energetic enough to compete
with the temperature and chemical potential and to allow adiabatic
manipulation on the sub-ms timescale. One way to achieve these
requirements is with intensity gradients of light, which make
neutral atom traps by virtue of the optical dipole force. Major
progress has been made with this approach
\cite{Bloch,Esslinger,Phillips,Ketterleinterferometer}, but still,
the light is not arbitrarily configurable and it is difficult to
address specific sites of an optical lattice. Structures
microfabricated on a surface, known as atom chips, are emerging as
a very promising alternative
\cite{Hindsreview,Schmiedmayerreview}. These can be patterned in
complex arrays on micrometer length scales. The locally-addressed
electric, magnetic and optical fields on a chip provide great
flexibility for manipulating and addressing the atoms. Magnetic
traps on atom chips are commonly generated either by
microfabricated current-carrying wires \cite{Schmiedmayerreview}
or by poled ferromagnetic films \cite{Hindsreview, Eriksson}
attached to some dielectric or metallic substrate. These are used
to create local minima of the magnetic field strength in which
low-field-seeking alkali atoms are trapped by the Zeeman effect.

In order to utilize atom chip structures of small scale, the atoms
must be held close to the surface. However, this same proximity
threatens to decohere the quantum state of the atoms through
electromagnetic field fluctuations that occur in the vicinity of a
surface. The free photon field does not perturb ground state
alkali atoms appreciably, but the evanescent field modes
associated with surface currents can be dense enough to generate
significant rf noise. This effect arises because the resistivity
of the surface material is always accompanied by field
fluctuations as a consequence of the fluctuation-dissipation
theorem. Several experimental studies have recently shown that rf
spin flip transitions occur when atoms are held close to thick
metallic or dielectric surfaces \cite{Jones,Cornell,Vuletic}.
Comparison with theory \cite{Henkel,Rekdal} has shown that this
spin relaxation is indeed due to thermal fluctuations of the
surface modes.

In this article, we explore the possibilities for reducing the
spin decoherence due to surface fields by making metallic surfaces
thin and by the possible use of superconducting materials.
Previous studies have found valuable scaling laws for the lifetime
near thick metallic slabs \cite{Henkel} and multi-layer wires
\cite{Rekdal}. The new results we derive here are of interest
because they describe the current generation of atom chips using
thin films and can guide future designs to achieve long qubit
coherence times.

Consider a ground-state alkali atom in hyperfine magnetic state
$|i\rangle$ and trapped at position $\textbf{r}_A$ near a surface.
The rate of the magnetic spin flip transition to state $|f\rangle$
has been derived by Rekdal \textit{et al.} \cite{Rekdal} as
\begin{eqnarray}
\label{eq:rate} \lefteqn{ \Gamma = \mu_0 \frac{2(\mu_B
g_S)^2}{\hbar}
\langle f|\hat{S}_j|i\rangle\langle
i|\hat{S}_k|f\rangle } \nonumber \\ && \textrm{Im}\left[
\bm{\nabla}\times\bm{\nabla}\times\bm{G}(\textbf{r}_A,\textbf{r}_A,
\omega) \right]_{jk} (\bar{n}_{\text{th}}+1) \,.
\end{eqnarray}
Here $\mu_B$ is the Bohr magneton, $g_S\approx 2$ is the electron
spin $g$-factor and $\langle f|\hat{S}_j|i\rangle$ is the matrix
element of the electron spin operator corresponding to the
transition $|i\rangle\mapsto|f\rangle$. Thermal excitations of the
electromagnetic field modes are accounted for by the factor
$(\bar{n}_{\text{th}}+1)$, where
\begin{equation}
\label{eq:bose}
\bar{n}_{\text{th}} = \frac{1}{e^{\hbar\omega/k_BT}-1}
\end{equation}
is the mean number of thermal photons per mode at the frequency
$\omega$ of the spin flip. The dyadic Green tensor
$\bm{G}(\textbf{r}_A,\textbf{r}_A,\omega)$ is the unique solution
to the Helmholtz equation
\begin{equation}
\bm{\nabla}\times\bm{\nabla}\times
\bm{G}(\textbf{r},\textbf{r}',\omega)
-\frac{\omega^2}{c^2}\varepsilon(\textbf{r},\omega)
\bm{G}(\textbf{r},\textbf{r}',\omega) =
\delta(\textbf{r}-\textbf{r}') \textbf{U},
\end{equation}
$\textbf{U}$ being the unit dyad. This tensor contains all
relevant information about the geometry of the set-up and also,
through the dielectric permittivity
$\varepsilon(\textbf{r},\omega)$, about the electric properties of
the surface. Equation~(\ref{eq:rate}) follows from a consistent
quantum-mechanical treatment of electromagnetic radiation in the
presence of absorbing bodies (for a review, see \cite{Buch}). It
is obtained by by considering the Heisenberg equations of motion
for a quantized magnetic dipole in the rotating-wave and Markov
approximations. The result is similar to calculations using
Fermi's Golden Rule \cite{Henkel}, where the local density of
states plays the r\^{o}le of the imaginary part of $\bm{G}$.

\begin{figure}[ht]
\includegraphics[width=5cm]{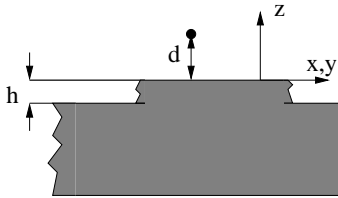}
\caption{\label{fig:slab} Schematic geometrical set-up. A plane
metallic film of thickness $h$ lies parallel to the $(x,z)$-plane
above a thick non-metallic substrate. The atom is located in
vacuum at a distance $d$ from the surface.}
\end{figure}
The geometry we are considering is illustrated in
Fig.~\ref{fig:slab}. We assume that a metallic slab of thickness
$h$ extends to infinity in the $x$ and $z$ directions (this is
solely for the computational simplicity that follows from
translational invariance in two directions). There is a thick
non-metallic substrate below and a vacuum above, where the atom is
located at a distance $d$ from the surface of the metal. Our
choice of $z$-axis corresponds to having a bias magnetic field
parallel to the surface, as is normally the case for a
Ioffe-Pritchard trap above an atom chip. The Green function for
this 3-layer structure, which is needed in order to use
Eq.~(\ref{eq:rate}), is commonly expressed in terms of a series of
cylindrical vector wave functions with appropriately chosen
generalized (Fresnel) reflection coefficients\cite{green}. There
are straightforward numerical routines that compute the required
elements of the Green tensor.
\begin{figure}[ht]
\includegraphics[width=8cm,angle=0,clip=]{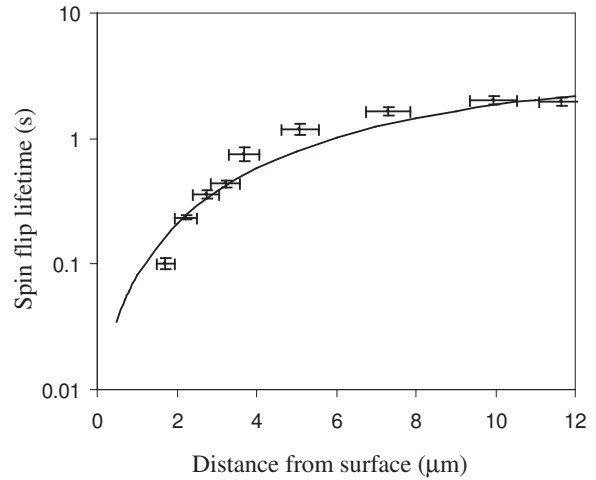}
\caption{\label{fig:distance} Lifetime $\tau$ as a function of
atom-surface distance $d$. Points: data given in
ref.\cite{Vuletic}. Line: Calculated lifetime using a skin depth
of $103\mu$m, a temperature of $400$K, and a frequency of $400$kHz
to coincide with the parameters of ref.\cite{Vuletic}. We also
include a factor of $5/3$ as discussed in ref.\cite{Vuletic} to
account for the two steps involved in spin-flip loss and we
include the loss due to background gas collisions.}
\end{figure}

Recent experiments measuring spin flip relaxation rates of atoms
trapped near thick surfaces have demonstrated the importance of
thermal field fluctuations \cite{Jones,Cornell,Vuletic}. This has
promoted great interest in thin surfaces because they should
generate less thermal noise, a conjecture that we confirm here. A
recent publication \cite{Vuletic} gives experimental values for
the loss rate of ${}^{87}$Rb atoms in the
$|5S_{1/2},F=2,m_F=2\rangle$ state, magnetically trapped near a
thin surface. The surface was a $2\mu$m-thick copper layer on a
substrate of nitride-coated silicon. The data points shown in
Fig.~\ref{fig:distance} reproduce the lifetimes for loss of atoms
from the trap in \cite{Vuletic}. At distances greater than about 7
$\mu$m from the surface, the loss rate is essentially constant and
is due to collisions with the background gas. At shorter
distances, the lifetime is reduced by thermally-induced spin
relaxation.  Seeking a comparison with theory, the authors
interpolated scaling laws given in \cite{Henkel} and found
agreement between theory and experiment for distances down to
$3.4\mu$m. Below that, there seemed to be a discrepancy, with the
observed lifetimes being substantially shorter than expected. It
was surmised that this discrepancy might be due to patch
potentials on the surface.

In the hope of resolving the discrepancy, we have calculated the
lifetimes from Eq.~(\ref{eq:rate}). This was done numerically,
using the Green's function technique discussed above. Our result
is shown as the solid line in Fig.~\ref{fig:distance}. For the
permittivity of the substrate, we ignored the silicon nitride and
took $\epsilon=11.7$ corresponding to the silicon, but the result
is not appreciably different even for $\epsilon=1$ because the
permittivity of the metal layer ($\sim10^{12}i$) is so high that
the Fresnel coefficients are not sensitive to such detail. At the
greatest distances $d$ in Fig.~\ref{fig:distance} there is just
the residual gas lifetime given by the authors of the experiment.
Below $10\mu$m, our calculation gives a slightly low lifetime
because the metal surface in the experiment was only $10\mu$m
wide, rather than being infinitely wide as our calculation
supposes. At lower heights still, where infinite width is a good
approximation, we again see agreement with the experiment. This
result indicates that the measurements in ref.\cite{Vuletic} were
correct and there is no need to invoke a possible contamination of
the surface.

The spin flip lifetime for the transition
$(F,m_F)=(2,2)\rightarrow(2,1)$ depends on three independent
length scales: the substrate thickness $h$, the atom-surface
distance $d$, and the skin depth $\delta$ of the substrate
material, defined via the Drude relation
$\varepsilon(\omega)\approx 2i(\frac{c}{\omega\delta})^2$
\cite{Jackson}. For certain regimes of these parameters it is
possible to approximate the integrals involved to obtain
analytical results for the lifetime $\tau=1/\Gamma$. Our results
are
\begin{equation}
\label{eq:behaviour} \tau \approx \left(\frac{8}{3}\right)^2
\frac{\tau_0}{\bar{n}_{\text{th}}+1}
\left(\frac{\omega}{c}\right)^3\left\{
\begin{array}{ll}
\frac{d^4}{3\delta} & \delta \ll d, h \\
\frac{\delta^2 d}{2} & \delta, h \gg d \\
\frac{\delta^2 d^2}{2h} &  \delta \gg d \gg h
\end{array}
\right. \,.
\end{equation}
Here, $\tau_0$ is the lifetime in free space at zero temperature,
given by $3\pi\hbar c^3/\mu_0\omega^3\sum|\langle
f|g_s\mu_B\hat{S}_j|i\rangle|^2$. At a transition frequency of
$\omega/2\pi=400$kHz this has the value $3\times 10^{25}$s. At a
temperature of $400$K, the factor $(\bar{n}_{\text{th}}+1)$
reduces the free-space lifetime to $4\times 10^{18}$s, but this is
still very long, being approximately the age of the universe. The
remaining factors take into account the effect of the surface and
these lead to much more dramatic reductions in lifetime. The first
two results in Eq.~(\ref{eq:behaviour}) describe the case of a
thick slab and are already known from ref. \cite{Henkel}. The
third result is new and describes the case of a thin film, which
is the case for most atom chips in use today.

\begin{figure}[ht]
\includegraphics[width=5cm,angle=-90]{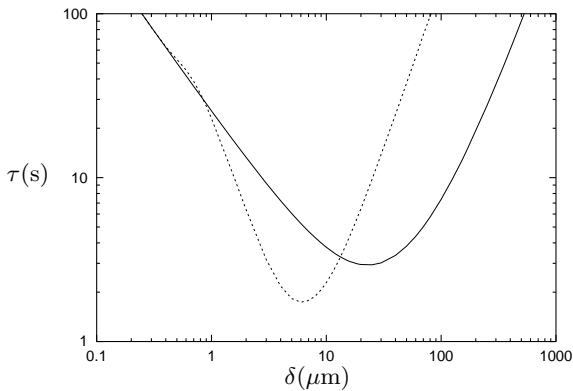}
\begin{picture}(0,0)
\put(-220,-70){$\tau$(s)}
\put(-115,-147){$\delta$($\mu$m)}
\end{picture}
\caption{\label{fig:delta} Lifetime $\tau$ as a function of skin
depth $\delta$ with the atom-surface distance fixed at $50\mu$m.
Solid line: Infinitely thick surface. Dotted line: $1\mu$m thick
surface.  We have taken a spin-flip frequency of $560$kHz and a
temperature of  $300$K.}
\end{figure}

In order to illustrate some aspects of these results,
Fig.~\ref{fig:delta} shows the spin-flip lifetime versus the skin
depth of the metal film for the same Rb transition in an atom
placed $50\mu$m away from the surface. The two curves correspond
to an infinitely thick film (solid line) and to a $1\mu$m-thick
film (dotted).
Where the skin depth is less than $1\mu$m, the two cases are
effectively the same because the source of the noise lies within
approximately one skin depth of the surface. Here the lifetime
scales as $\delta^{-1}$ in accordance with the first line of
Eq.~(\ref{eq:behaviour}). As the skin depth becomes longer we
enter a range where $h \ll \delta \ll d$ for the thin film. Here
the thin film produces a shorter lifetime than the thick one,
somewhat surprisingly. Once $\delta$ becomes large compared with
$d$, the case of the thick film is described by the second line of
Eq.~(\ref{eq:behaviour}) whereas that of the thin film follows the
third line.  In either case $\tau\propto\delta^2$, as can be seen
on the right side of Fig.~\ref{fig:delta}, with the important
difference that the thin film gives a longer lifetime by a factor
of $d/h$.

Between the large and small extremes of skin depth the lifetime
exhibits a minimum (see also \cite{Rekdal}).  For thick films, we
find the minimum at $\delta_{\text{min}}\simeq d$, whereas for
thin films it is at $\delta_{\text{min}}\simeq \sqrt{h d}$.
Evidently the minimum represents a condition for coupling the
excitation most efficiently out of the atom and into surface
excitations - a kind of impedance matching. One consequence of the
minimum is that for any fixed atom-surface distance $d$, there are
two possible choices for the skin depth of the metallic film to
produce a given lifetime. For example, with the atom placed
$50\mu$m above a thick slab, Fig.~\ref{fig:delta} shows that skin
depths of $1\mu$m and $100\mu$m both lead to a $10$s lifetime. At
the $560$kHz frequency used for this figure, the larger skin depth
corresponds to a slab of metal such as Cu ($\delta=85\mu$m) or Al
($\delta=110\mu$m), both excellent conductors.

There are of course no normal metals with a skin depth at $560$kHz
as small as $1\mu$m (a resistivity of $2\times
10^{-12}\Omega$\,m), but superconductors are possible candidates.
In a material with
 superconducting gap $\Delta(T)$ at temperature $T$, the usual
Maxwell-Boltzmann distribution $\text{exp}(-2\Delta(T)/k_BT)$
determines the fraction of Cooper pairs that are thermally broken
to form a gas of normally conducting electrons \cite{Tinkham}.
Typically $\Delta(0)\simeq k_BT_c$, where $T_c$ is the transition
temperature.  Thus, at temperatures that are moderately below $
T_c$ there is a significant fraction of normally-conducting
electrons. On the other hand, when $T\ll T_c$, this fraction
becomes vanishingly small.

One particularly relevant superconducting material for possible
use in atom chips is niobium, because it has a high transition
temperature. In bulk material $T_c=9.3$K \cite{Klein}, while
$T_c=8.3$K has been measured for films with $15$nm thickness
\cite{Pronin}. The superconducting energy gap is estimated to be
$\Delta(0)\approx 2.1k_BT_c$ \cite{Pronin}. Measurements of the
complex magnetic susceptibility of ultra-pure niobium (residual
resistivity ratio $RRR=300$) have recently been published in
\cite{Casalbuoni}. These are of particular interest here because
they provide explicit figures for the real part of the complex
conductivity $\sigma(\omega)$ at frequencies $\leq 1$MHz. Just
above the superconducting transition temperature the conductivity
is $2\times 10^9 (\Omega \text{m})^{-1}$ \cite{Casalbuoni}, which,
through the relation $\delta^2=2/(\mu_0 \omega\sigma)$, gives a
skin depth in the normal state at $560$kHz of
$\delta_{\text{N}}\simeq 15\mu$m. The magnetic susceptibility
measurements of \cite{Casalbuoni} show a hundredfold increase in
conductivity when the temperature drops to $T\simeq 4$K,
corresponding to a skin depth of $1\text{--}2\mu$m. This is
significantly larger than the zero-temperature London penetration
depth of $46\pm 2$nm \cite{Casalbuoni}.

This analysis shows that i) superconducting films have the
potential to provide surfaces with skin depths of 1 micron or
less.  ii) that the atom-surface distances similar in magnitude to
the skin depth are to be avoided. For atom chips with the atoms at
tens of microns away from the surface, the use of superconducting
niobium wires at 4K can boost the spin relaxation time to
$~10^3$s. This boost comes partly from the lower temperature which
accounts for a 100 times smaller value of
$\bar{n}_{\text{th}}$. This enhancement would be present for normal
metals as well. From Fig.~\ref{fig:delta} we also see that part of
this boost comes from the smaller skin depth of superconductors.
However, small scale trapping structures are required for many
quantum information processing schemes (e.g. \cite{Calarco}), and
then it is natural to hold the atoms one or two microns away from
the surface. In these cases, the unfortunate similarity between
the atom-surface distance and the skin depth can make a
superconducting surface a worse choice than a normal metal.

In conclusion, we have used a consistent quantum-theoretical
description of electromagnetic radiation near metallic/dielectric
bodies to derive an expression for the spin relaxation lifetime of
a neutral atom held near the thin plane metallic surface of an
atom chip. We have been able to show that the lifetime reported
near such an atom chip by the group of Vuleti\'{c} \cite{Vuletic}
is consistent with this theory. We have found that the
spin-relaxation lifetime of an atom trapped at a given height
above a metallic surface exhibits a minimum with respect to the
skin depth of the surface. For atoms placed tens or hundreds of
microns away from the surface, superconducting atom chips at low
temperature offer improved lifetimes. However, we find that when
atoms are placed only a few microns from the surface, as in many
current atom chip experiments, the spin relaxation above normal
metals is liable to be slower than above a superconductor. These
results will be helpful in guiding the design of future
miniaturized atom chips.

We acknowledge valuable discussions with D. Lee.  This work was
supported by the UK Engineering and Physical Sciences Research
Council (EPSRC) and the European Commission (FASTnet and QGATES
programmes). P.K.R. acknowledges support by the Research Council
of Norway. S.S. enjoys the support of the EPSRC through an
Advanced Research Fellowship.


\end{document}